\documentclass{article}
\usepackage{amsmath}
\usepackage{amsthm}
\usepackage{amssymb}
\usepackage{bm}
\usepackage[hmargin=3cm,vmargin=3.5cm]{geometry}

\newtheorem{theorem}{Theorem}
\begin{document}
%
%
%
\title{An efficient algorithm for sampling from $\sin^k(x)$ for generating random correlation matrices}
\author{Enes Makalic \and Daniel F. Schmidt}
\maketitle
\begin{abstract}
In this note, we develop a novel algorithm for generating random numbers from a distribution with a probability density function proportional to $\sin^k(x)$, $x \in (0,\pi)$ and $k \geq 1$. Our algorithm is highly efficient and is based on rejection sampling where the envelope distribution is an appropriately chosen beta distribution. An example application illustrating how the new algorithm can be used to generate random correlation matrices is discussed.
\end{abstract}
\section{Introduction}
Consider a positive random variable $X$ with probability density function 
\begin{equation}
	\label{eqn:sin_x_k}
	f_X(x) = c_k \sin^k(x), \quad x \in (0,\pi), \quad k \geq 1,
\end{equation}
where the normalization constant $c_k$ is given by
\[
	c_k = \frac{\Gamma \left(\frac{k}{2}+1\right)}{\sqrt{\pi } \Gamma \left(\frac{k}{2}+\frac{1}{2}\right)} \,.
\]
This is a symmetric continuous probability density function for which the mean (mode and median) are
\[
	{\rm E}_X[X] = \pi/2, \quad f_X(\pi/2) = c_k \, ,
\]
for all $k \geq 1$. Generating random numbers from this distribution is not straightforward as the corresponding cumulative density function, although available in closed form, is defined recursively and requires $O(k)$ operations to evaluate. The nature of the cumulative density function makes any procedure based on inverse transform sampling computationally inefficient, especially for large $k$. Furthermore, an alternative approach based on numerical integration is not recommended due to numerical issues with the probability density function when $k$ is large.

In this note, we develop a novel algorithm (see Section~\ref{sec:algorithm}) for generating random numbers from the distribution defined in (\ref{eqn:sin_x_k}). Our algorithm is based on rejection sampling~\cite{RobertCasella04} (pp. 50--53) where the envelope distribution is a particular beta distribution. We compute the exact unconditional acceptance probability of the proposed algorithm and show that the maximum expected average number of iterations per sample is $\pi/(2\sqrt{2})$ $(\approx 1.11)$ for all $x \in (0, \pi)$ and $k \geq 1$, which highlights the efficiency of the proposed algorithm. Lastly, we use our novel rejection sampler to implement the algorithm for generating high dimensional correlation matrices based on the hyperspherical parametarization (see Section~\ref{sec:application}) proposed by Pourahmadi and Wang~\cite{PourahmadiWang15}. Our implementation of the correlation matrix sampling algorithm for the Matlab and R numerical computing platforms is available for download from Matlab Central (File ID: 68810) and CRAN (the {\tt randcorr} package).
\section{Rejection sampling algorithm}
\label{sec:algorithm}
The proposed algorithm for sampling from density ({\ref{eqn:sin_x_k}) utilises rejection sampling where the envelope distribution is an appropriately scaled symmetric ${\rm Beta}(k+1,k+1)$ distribution with probability density function 
\begin{equation}
	\label{eqn:beta}
	g_Y(x) = \frac{ x^k (\pi - x)^k} {{\rm B}(k+1,k+1) \pi^{2k+1}}, \quad x \in (0,\pi), \quad k \geq 1,
\end{equation}
where ${\rm B}(\cdot,\cdot)$ is the Euler beta function. Theorem~\ref{thm:main} proves that $g_Y(y)$ is an envelope density for this problem and gives the expected average sampling efficiency for all $k \geq 1$. The proof of the theorem is given in Appendix~\ref{sec:proof}.

\begin{theorem}
\label{thm:main}
Let $f_X(x)$ denote the probability density function given in (\ref{eqn:sin_x_k}) and let $Y$ denote a ${\rm Beta}(k+1,k+1)$ random variable with probability density function $g_Y(x)$ (\ref{eqn:beta}). Then,
\begin{equation}
	f_X(x) \leq M_k \, g_Y(x), \quad x \in (0,\pi), \quad k \geq 1,
\end{equation}
where $M_k$ is the upper-bound of the likelihood ratio $f_X(x) / g_Y(x)$ and is given by
\begin{equation}
	\label{eqn:Mk}
	M_k = \frac{f_X( \pi/2 )}{g_Y(\pi / 2)} = \frac{\sqrt{\pi } 2^{k-1} \Gamma \left(\frac{k}{2}+1\right)^2}{\Gamma \left(k+\frac{3}{2}\right)} \, .
\end{equation}
\end{theorem}

Based on Theorem~\ref{thm:main}, our algorithm to generate a random variable $X$ with probability density function (\ref{eqn:sin_x_k}) is given below:
\begin{enumerate}
	\item Generate $X \sim {\rm Beta}(k+1,k+1)$ 
	\item Generate $U$ from the uniform distribution $U \sim \mathcal{U}_{[0,1]}$
	\item Accept $X$ if 
	\[
		(1/k) \log U \leq \log \left( \frac{\pi^2 \sin X}{4 X(\pi-X)} \right)
	\]
	\item Otherwise, return to Step 1.
\end{enumerate}

The average number of iterations required to obtain a sample from the distribution (\ref{eqn:sin_x_k}) is given by $M_k$ (\ref{eqn:Mk}). This is a strictly increasing function of $k \geq 1$ where the first few values are
\[
	M_1 = \frac{\pi}{3}, \quad M_2 = \frac{16}{15}, \quad M_3 = \frac{12 \pi }{35}, \quad M_4 = \frac{1024}{945} \, .
\]
The maximum of $M_k$ is given by
\[
	\lim_{k \to \infty} M_k = \frac{\pi }{2 \sqrt{2}} \approx 1.11
\]
which emphasises the efficiency of the proposed algorithm. In particular, for any value of $k \geq 1$, the maximum expected average number of iterations per sample is $\pi/(2\sqrt{2}) \approx 1.11$.

\section{Application}
\label{sec:application}
The proposed rejection sampling algorithm is suitable for applications requiring repeated sampling from density (\ref{eqn:sin_x_k}), especially for large $k$. An example application is the problem of random sampling of a $p \times p$ correlation matrix ${\bf R}$, which may be high dimensional. For this problem, we implement the procedure outlined in~\cite{PourahmadiWang15} which consists of two steps: (i)~generate $p(p-1)/2$ angles $0 < \theta_{ij} < \pi$ from the distribution (\ref{eqn:sin_x_k}) where $k = \{1,\ldots,p-1\}$; and (ii)~compute a lower triangular matrix ${\bf B}$ (equation (1) in~\cite{PourahmadiWang15}) from the angles $\theta_{ij}$ and then the corresponding $p \times p$ correlation matrix ${\bf R} = {\bf B} {\bf B}^{\rm T}$ (see~\cite{PourahmadiWang15} for details). 

The procedure of Pourahmadi and Wang requires a computationally efficient method for sampling of the angles $\theta_{ij}$. This is especially true for high dimensional correlation matrices where $p$ (and therefore $k$) is large and the number of random variates required is of the order $O(p^2)$. Using our rejection sampling algorithm, we have implemented the Pourahmadi and Wang sampler in the Matlab and R  numerical computing environments. Our program can generate a random $100 \times 100$ correlation matrix (i.e., 4,950 samples of $\theta_{ij}$) in $\approx$0.02s, and a random $1,000 \times 1,000$ (i.e., 499,500 samples of $\theta_{ij}$) correlation matrix in $\approx$0.5s on a commodity laptop. In contrast, the inverse transform sampling approach implemented in~\cite{PourahmadiWang15} and running on comparable computing hardware, requires approximately $0.9s$ and $42s$ to generate a $100\times 100$ and a $600 \times 600$ correlation matrix, respectively. 

\appendix
\section{Proof of Theorem~\ref{thm:main}}
\label{sec:proof}
\begin{proof}{}
We wish to show that $f_X(x) \leq M_k g_Y(x)$, for all $x \in (0,\pi)$ and $k \geq 1$. First, note that
\begin{eqnarray}
	f_X(x) &\leq& M_k \, g_Y(x) \nonumber \\
	\frac{\Gamma \left(\frac{k}{2}+1\right) \sin^k(x)}{\sqrt{\pi } \Gamma \left(\frac{k}{2}+\frac{1}{2}\right)} &\leq& \left( \frac{\sqrt{\pi } 2^{k-1} \Gamma \left(\frac{k}{2}+1\right)^2}{\Gamma \left(k+\frac{3}{2}\right)}\right)  \frac{ x^k (\pi - x)^k} {{\rm B}(k+1,k+1) \pi^{2k+1}} \nonumber \\
	\sin^k(x) &\leq& 4^k \pi ^{-2 k} ((\pi -x) x)^k \nonumber \\
	\sin(x) &\leq& 4 \pi^{-2} (\pi -x) x \nonumber 
\end{eqnarray}
The function $p_3(x) = 4 \pi^{-2} (\pi -x) x$ is an interpolation polynomial at the nodes $x_0=0,x_1=x_2=\pi/2$ and $x_3=\pi$ for the function $\sin(x)$ for all $x \in (0,\pi)$~\cite{MartinR16}. The interpolation error for the polynomial $p_3(x)$ and some $\xi \in (0,\pi)$ is
\begin{eqnarray}
	\sin(x) - p_3(x) = \frac{\sin(\xi)}{24} \prod_{j=0}^3 (x - x_j) = \frac{x}{24} \left(x-\frac{\pi }{2}\right)^2 (x-\pi ) \sin (\xi ) \leq 0 \, , \nonumber
\end{eqnarray}
as the only term that is negative is $(x - \pi) \leq 0$. The inequality is strict everywhere except the boundary points ($x = 0, \pi$) and the maximum of $f_X(x)$ and $g_Y(x)$ ($x = \pi/2$). 
\end{proof}

\bibliography{R:/5050/CEB/Share/Staff/EnesAndDaniel/Bibliography/bibliography} 

\begin{thebibliography}{1}
\expandafter\ifx\csname url\endcsname\relax
  \def\url#1{\texttt{#1}}\fi
\expandafter\ifx\csname urlprefix\endcsname\relax\def\urlprefix{URL }\fi
\expandafter\ifx\csname href\endcsname\relax
  \def\href#1#2{#2} \def\path#1{#1}\fi

\bibitem{RobertCasella04}
C.~P. Robert, G.~Casella, Monte {C}arlo Statistical Methods, Springer, 2004.

\bibitem{PourahmadiWang15}
M.~Pourahmadi, X.~Wang, Distribution of random correlation matrices:
  Hyperspherical parameterization of the {C}holesky factor, Statistics \&
  Probability Letters 106 (2015) 5--12.

\bibitem{MartinR16}
M.~R. (user:42969), How to prove $\sin(x) \leq 4 x (\pi - x) / \pi^2$ for all
  $x \in (0,\pi)$?, Mathematics, https://math.stackexchange.com/q/2030069
  (2016).

\end{thebibliography}

\end{document}